\documentclass[superscriptaddress,prl,twocolumn,nofootinbib]{revtex4}
\usepackage{amsmath,amssymb}
\usepackage{graphicx,epsfig}
\usepackage{slashed}
\usepackage{float}

\def\beq{\begin{equation}}
\def\eeq#1{\label{#1}\end{equation}}
\def\eeqn{\end{equation}}
\def\beqa{\begin{eqnarray}}
\def\eeqa#1{\label{#1}\end{eqnarray}}
\def\eeqan{\end{eqnarray}}

\def\leqn#1{(\ref{#1})}
\newcommand{\bspace}{\!\!\!\!}

\def\met{\mbox{$E{\bspace}/_{T}$}}

\newcommand{\centeron}[2]{{\setbox0=\hbox{#1}\setbox1=\hbox{#2}\ifdim
\wd1>\wd0\kern.5\wd1\kern-.5\wd0\fi \copy0
\kern-.5\wd0\kern-.5\wd1\copy1\ifdim\wd0>\wd1
                                   \kern.5\wd0\kern-.5\wd1\fi}}
\newcommand{\ltap}{\>\centeron{\raise.35ex\hbox{$<$}}
                           {\lower.65ex\hbox{$\sim$}}\>}
\newcommand{\gtap}{\>\centeron{\raise.35ex\hbox{$>$}}
                           {\lower.65ex\hbox{$\sim$}}\>}
\newcommand{\gsim}{\mathrel{\gtap}}

\begin{document}

\title{Boosted Tops from Gluino Decays}

\author{Joshua Berger}
\affiliation{Laboratory for Elementary Particle Physics, Cornell University,
Ithaca, NY 14850, USA}
\author{Maxim Perelstein}
\affiliation{Laboratory for Elementary Particle Physics, Cornell University,
Ithaca, NY 14850, USA}
\author{Michael Saelim}
\affiliation{Laboratory for Elementary Particle Physics, Cornell University,
Ithaca, NY 14850, USA}
\author{Andrew Spray}
\affiliation{TRIUMF, 4004 Wesbrook Mall, Vancouver, BC V6T 2A3, Canada}

%\date{\today}

\begin{abstract}
Naturalness considerations, together with the non-observation of superpartners of the Standard Model particles at the Large Hadron Collider (LHC) so far, favor supersymmetric (SUSY) models in which third generation squarks are significantly lighter than those of the first two generations. In such models, gluino pair-production is typically the dominant SUSY production process at the LHC, and it often leads to final states with multiple top quarks. Some of these top quarks may be relativistic in the lab frame, in which case their hadronic decays may produce ``top jets". We propose that the recently developed techniques for tagging top jets can be used to boost sensitivity of the LHC searches for this scenario. For example, within the simplified model used for this study, we estimate that a search with 2 top-tagged jets can probe gluino masses of up to about 1 TeV at the 7 TeV LHC with 30 fb$^{-1}$ integrated luminosity.   
\end{abstract}


\maketitle
{\em Introduction ---} Recently, experiments at the Large Hadron Collider (LHC) have begun searching for new physics beyond the Standard Model (SM). Among the many theoretical ideas about the possible nature of this new physics, supersymmetry (SUSY) is the most popular one: it provides an appealing solution to the gauge hierarchy problem of the SM, contains an attractive dark matter candidate, and fits naturally in the framework of grand unification and string theory. SUSY models predict a number of new particles, ``superpartners" of the known SM particles, which may be produced at the LHC. In the simplest SUSY models, all superpartners are odd under a discrete symmetry, R-parity, while all SM particles are R-even. This implies that the lightest SUSY particle (LSP) is stable, and that any other superpartner will decay to the LSP and one or more SM particles. Cosmological considerations strongly prefer the LSP to be electrically neutral and uncolored, so that at the LHC the LSP passes through the detector without interactions, leading to an apparent transverse momentum imbalance, or ``missing transverse energy" (MET). The presence of MET provides a distinct signature which can be used to distinguish SUSY events from the (far more numerous) SM backgrounds.   

At the time of writing, the LHC experiments have presented searches for events with anomalous MET using a data set of approximately 1 fb$^{-1}$ collected in 2010-11 at the center-of-mass energy of $\sqrt{s}=7$ TeV. No evidence for anomalous MET has been found, and limits on superpartner masses have been set. Barring accidental features such as spectrum degeneracies, gluinos $\tilde{g}$ and squarks of the first two generations $\tilde{q}_{1,2}$ have been ruled out for masses up to about 1 TeV~\cite{LHC_data}. In models where all squarks have a common mass at some energy scale, this bound implies that a significant amount of fine-tuning would be necessary to accommodate the observed electroweak symmetry breaking scale~\cite{strumia}. On the other hand, fine-tuning can be avoided if the third-generation squarks, stops $\tilde{t}$ and sbottoms $\tilde{b}$, are significantly lighter than $\tilde{q}_{1,2}$~\cite{NSUSY1,NSUSY2,NSUSY3}. The LHC bounds on third-generation squarks are quite weak: stops above 200-300 GeV are currently allowed. The only other superpartner whose mass is significantly constrained by naturalness is the gluino~\cite{NSUSY1}; at present, gluinos above 600 GeV are allowed if decaying only via the 3rd generation. With this motivation, we will focus on a scenario where gluinos, third-generation squarks, and a neutralino LSP are  the only particles relevant for the LHC phenomenology, with other squarks being too heavy to be produced. An explicit example of a complete theory realizing this spectrum is the ``accidental SUSY" models of Refs.~\cite{Gnew}.

The lack of discovery so far also implies that traditional SUSY searches using the MET signature will become more difficult, since the large-MET tails of SM backgrounds will need to be calculated (or extrapolated) with increasingly high precision to obtain sensitivity to lower SUSY cross sections. This motivates the question: Can any handles other than MET be used to identify SUSY events in the presence of large SM backgrounds? In this Letter, we explore an alternative signature. Gluino cascade decays to the LSP via intermediate stops produce two top quarks, so that gluino pair-production events may result in final states with four tops~\cite{TW,Kane_tops}. If the gluino-stop and stop-LSP mass differences are sufficiently large, each of these tops will typically be relativistic in the lab frame, and its hadronic decay products will be merged into a single jet. Recently, much work has been done on distinguishing such top jets from the usual hadronic jets using the energy distribution inside the jet, and several well-tested algorithms for ``tagging" top jets are now available~\cite{boost2010}. The original motivation was to search for decays of the Kaluza-Klein gluon in models with extra dimensions~\cite{KKgluon}; other proposed applications include a search for the string-Regge excitation of the gluon~\cite{Regge}, and a search for direct stop production in SUSY~\cite{stops}. Here, we point out that this technique can also be used to search for the SUSY gluino, and is particularly promising in scenarios with a light third generation, since $\tilde{g}$ decays to tops have large branching fractions in this case. 

{\em Analysis Setup ---} In the spirit of the ``simplified model" approach~\cite{SimpM,Simp3}, we assume that a gluino $\tilde{g}$, one stop $\tilde{t}$, and a single neutralino $\tilde{\chi}^0$ are the only superpartners relevant for the LHC phenomenology. This is the minimal set of particles required to produce our signature. In the Minimal Supersymmetric Standard Model (MSSM), this setup can be realized if the second stop and the left-handed sbottom are heavier than the gluino. (Note that naturalness considerations in the MSSM prefer spectra with a few hundred-GeV splitting among the two stop mass eigenstates~\cite{golden}.) If this is not the case, the branching ratios of the decays producing our signature would be reduced (e.g. from 1 to 2/3 if all three squarks are degenerate), resulting in a somewhat decreased rate, but qualitatively the picture is unchanged. We assume that the neutralino is the stable LSP, and set its mass to 60 GeV throughout the analysis. The LHC signal is dominated by gluino pair-production, followed by the cascade decay
\beq
\tilde{g}\to \tilde{t} + \bar{t},~~~~\tilde{t} \to t \tilde{\chi}^0\,,
\eeq{decays} 
or its charge conjugate. We assume that $m(\tilde{g})-m(\tilde{t})>m_t$, $m(\tilde{t})-m(\tilde{\chi}^0)>m_t$, 
so that all four tops in the event are on-shell. (It may be possible to relax one of these conditions, as long as the other one is satisfied strongly so that at least two tops in the event are boosted; we will not study that possibility here.) We compute gluino pair-production cross sections at next-to-leading order (NLO) using {\tt PROSPINO}~\cite{prospino}. To study cut efficiencies, we generate event samples for gluino pair-production followed by the decays~\leqn{decays} using {\tt MadGraph/MadEvent v5 1.3.27 (MG/ME)}~\cite{MGME} for a large set of parameters $(m(\tilde{g}), m(\tilde{t}))$. We then simulate top decays, showering and hadronization with {\tt PYTHIA 8}~\cite{pythia}. To identify jets, we use the anti-$k_T$ algorithm implemented in the {\tt FastJet} code~\cite{FastJet,FastJetWeb}. Top tagging of jets in our sample is simulated using the implementation of the Hopkins algorithm~\cite{Hopkins} available at~\cite{FastJetWeb}. In the top tagger, we use two sets of parameters, ``tight" and ``loose" tags; they are defined precisely as in Ref.~\cite{boost2010}. 

We require at least 4 jets with $p_T>100$ GeV in each event, and require that some of the jets be top-tagged. (The optimal number of top-tagged jets required depends on the LHC energy and luminosity, see below.) In the signal, tagged jets are typically due to hadronic decays of boosted tops, which produce 3 collimated partons that cannot be resolved. The backgrounds include SM processes with boosted tops, as well as ordinary jets mistakenly tagged as top-jets. (The mistag probability is typically of order 1\%~\cite{boost2010}.) We also require the presence of substantial missing energy. The irreducible backgrounds may contain MET from invisible $Z$ decays, leptonic $W$ decays, or semileptonic top decays. We include the following irreducible backgrounds: $n t + (4-n) j$ with $n=1\ldots 4$; $Z+n t + (4-n) j$, with $n=0, 2, 4$; and $W+n t + (4-n) j$, with $n=0, 2, 4$. Here each $t$ may be a top or an anti-top, $j$ denotes a jet due to a non-top quark or a gluon, and $Z\to\nu\bar{\nu}$ or $W\to \ell \nu$ is required. We do not include reducible backgrounds, other than the light jets mistagged as tops. We simulated the backgrounds at parton level with {\tt MG/ME}, and used these samples to compute $p_T$ and MET cut efficiencies. We use leading-order (LO) cross sections for all background processes. The two dominant backgrounds, $2t+2j$ and $Z+4j$, have been recently computed at NLO. In both cases, the NLO correction to the cross section is negative: K-factors of 0.73 for $2t+2j$~\cite{2t2j} and 0.95 for $Z+4j$~\cite{z4j} have been reported, so that using LO cross sections for these processes is conservative. No other backgrounds are currently known beyond the LO.  

Unfortunately, due to large QCD rates and small mistag probabilities, we were not able to generate Monte Carlo samples large enough to measure top-tag efficiencies directly in the background channels. Instead, we estimate these efficiencies by multiplying the $p_T$-dependent tag and mistag probabilities for individual top and non-top jets reported in Ref.~\cite{boost2010}. This estimate assumes that the tag and mistag probabilities for each jet are independent of the presence of other objects in the final state (the probabilities in~\cite{boost2010} were computed using $t\bar{t}$ and $2j$ samples). The probability to tag a true top jet as such is clearly reduced by the presence of other jets in the event: for example, the tag efficiency for our signal approximated in this way is typically about a factor of two higher than that obtained by a full simulation. So, our estimate of backgrounds involving tops, such as $2t+2j$, is certainly conservative. It is less clear how the mis-tag probability would be affected; we leave this issue for future work. 

\begin{figure}[t]
\begin{center}
\includegraphics[width=9cm]{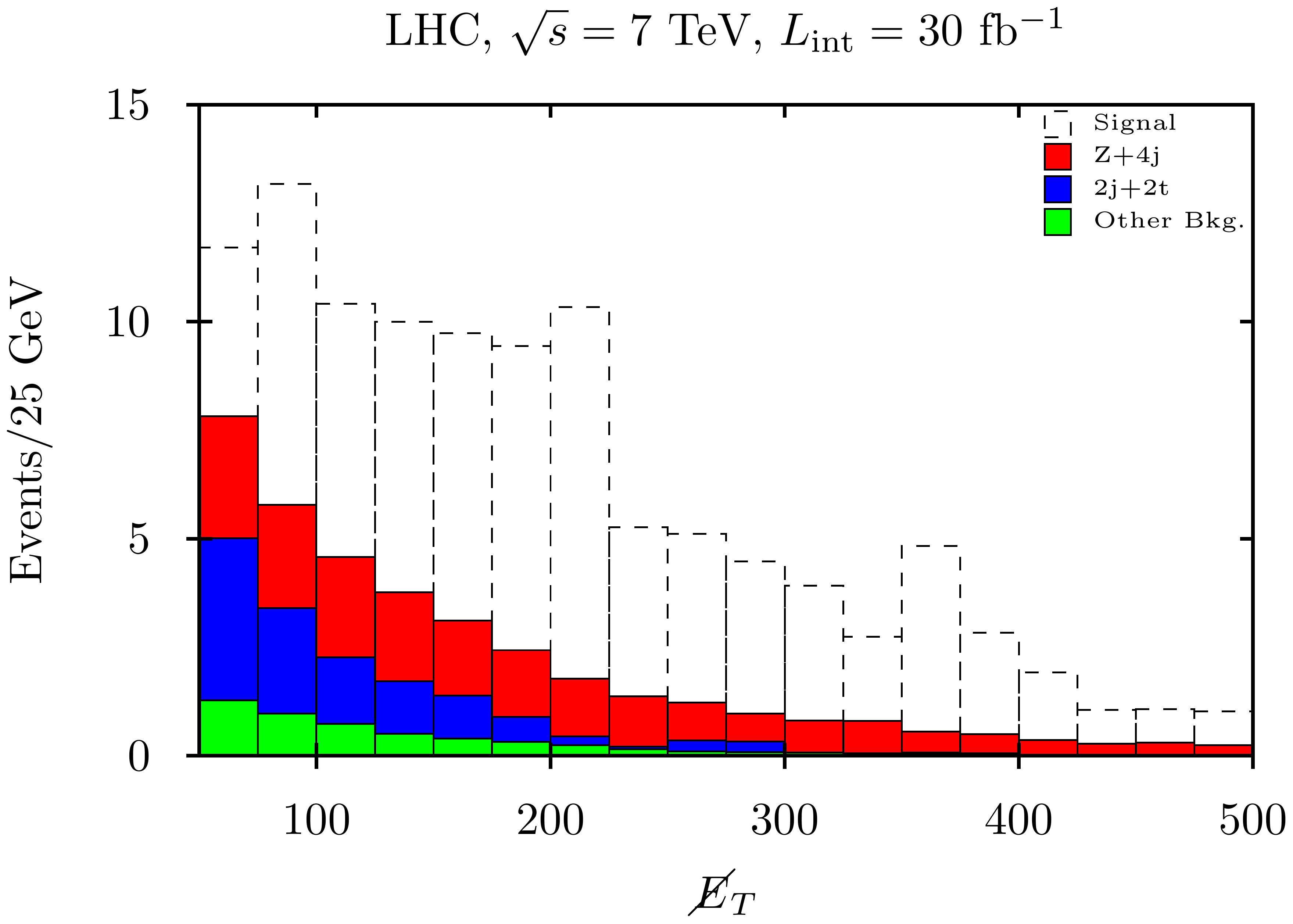}
\vspace{-8mm}
\caption{Signal at the benchmark point, $(m(\tilde{g}), m(\tilde{t}))=(800, 400)$ GeV, and background rates as a function of MET, at 7 TeV LHC. Four jets with $p_T>100$ GeV and two top-tagged jets are required.}
\label{fig:METhisto}
\end{center}
\end{figure}

\begin{figure}[h]
\begin{center}
\includegraphics[width=8cm]{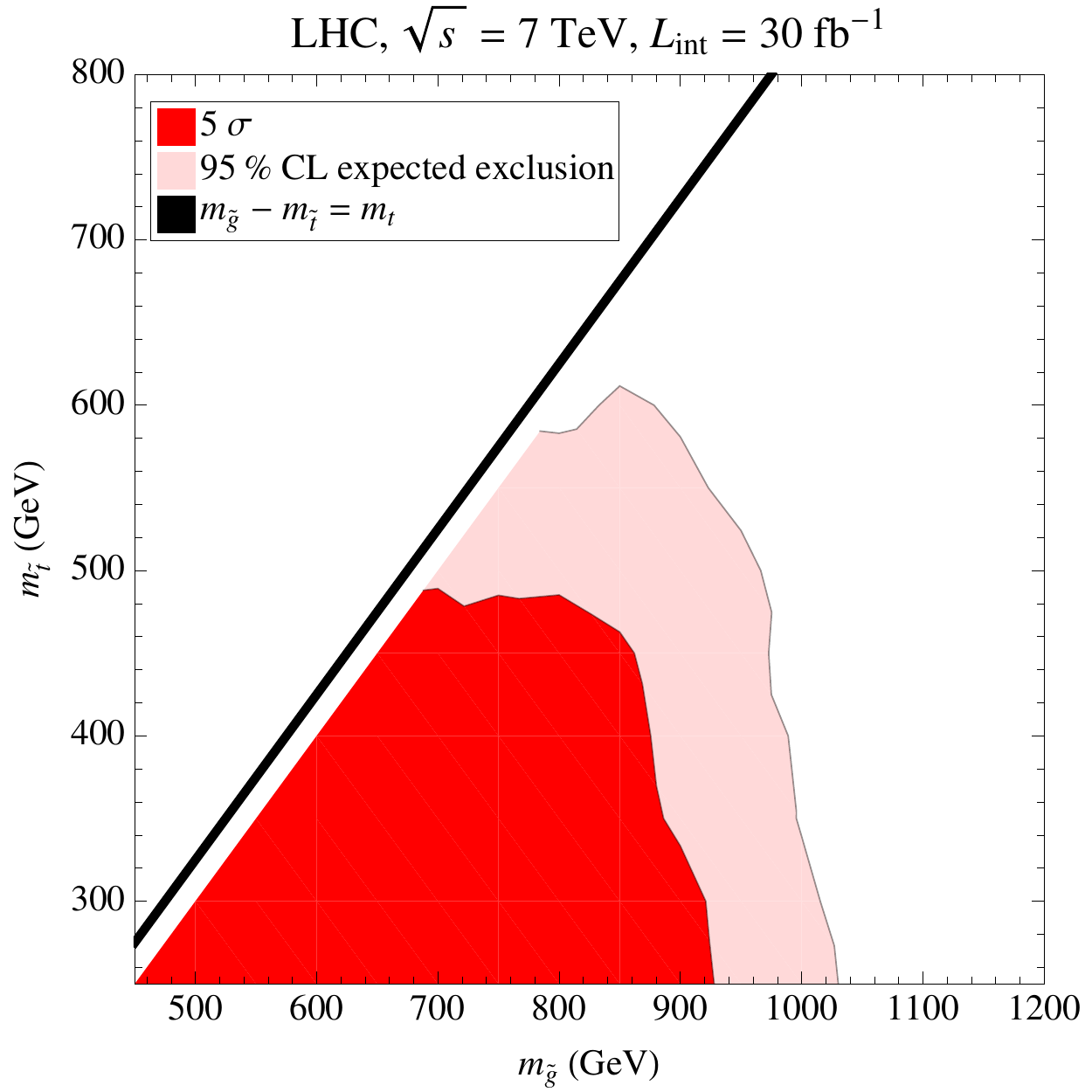}
\caption{The 95\% c.l. expected exclusion and 5-sigma discovery reach of the proposed search at the 7 TeV LHC run with 30 fb$^{-1}$ integrated luminosity.}
\label{fig:reach7}
\end{center}
%\vspace{-8mm}
\end{figure}

{\em LHC Sensitivity at $\sqrt{s}=7$ TeV ---} To keep the analysis simple, we optimize the selection cuts for 
a single ``benchmark" point in the model parameter space, and do not vary them as we scan the masses. At 7 TeV, we choose the benchmark point $(m(\tilde{g}), m(\tilde{t}))=(800, 400)$ GeV. We studied all possible combinations of between 0 and 4 loose and tight top tags, and conclude that requiring 2 loose tags is the best strategy at this point. Analyses requiring more than 2 tags, or 2 or more tight tags, suffer from low event rate, making a search in the 7 TeV LHC run with $20-30$ fb$^{-1}$ integrated luminosity impractical. Requiring fewer tags leads to significantly higher background rates, decreasing sensitivity~\cite{BB}. 
The two top tag requirements strongly suppress the backgrounds, as illustrated in Table~\ref{tab:cutflow}, but are not by themselves sufficient, so that an additional MET cut must be applied. The signal and principal backgrounds as a function of MET are shown in Fig.~\ref{fig:METhisto}. We require $\met>100$ GeV; with this cut, we expect $32$ signal events, $S/B=2.4$, and statistical significance of $6.8$ at the benchmark point with 30 fb$^{-1}$ integrated luminosity. The reach of the LHC with this data set is shown in Fig.~\ref{fig:reach7}. (The 95\% exclusion contour is calculated using the expected $CL_s$~\cite{Junk}.  The discovery significance is determined using the expected log likelihood of consistency with the signal plus background hypothesis~\cite{FC}.) Gluino masses of up to about $1$ TeV can be probed at the 95\% confidence level, as long as the gluino-stop mass difference exceeds $400$ GeV. The 5-sigma discovery reach extends to a gluino mass of about 900 GeV for stop masses below 350 GeV. We should also note that $S/B\gsim1$ throughout the probed region, so no extraordinarily precise predictions of the background are required. 

\begin{table}
\begin{tabular}{|l|c|c|c|c|c|c|}
\hline Process & $\sigma_{\rm tot}$ & Eff($p_T$) & Eff(tag) & $\sigma_{\rm tag}$ & Eff($\met$) & $\sigma_{\rm all~cuts}$ \\
\hline
signal & 61.5 & 37 & 6 & 1.31 & 81 & 1.06 \\
\hline
$Z+4j$ & $2 \times 10^5$ & 0.2 & 0.1 & 0.44 & 66 & 0.29 \\
$2t+2j$ & $5 \times 10^4$ & 3 & 0.3 & 5.7 & 2 & 0.10 \\
$W+4j$ & $2\times 10^5$ & 0.2 & 0.03 & 0.12 & 29 & 0.04 \\
$Z+2t+2j$ & 50 & 4 & 1 & 0.02 & 72 & 0.02 \\ 
\hline
\end{tabular}
\caption{Signal and background cross sections (in fb) and cut efficiencies (in \%) at the 7 TeV LHC. Acceptance cuts of $p_T>20$ GeV, $|\eta|<5$ for all jets are included in the total cross sections. The cuts are labelled as follows: ``$p_T$": requiring 4 jets with $p_T>100$ GeV; ``tag": requiring 2 jets to be tagged as tops with ``loose" parameters; ``$\met$": requiring $\met>100$ GeV. The signal is at the benchmark point, $(m(\tilde{g}), m(\tilde{t}))=(800, 400)$ GeV. Backgrounds not listed here are negligible.}
\label{tab:cutflow}
%\vspace{-4mm}
\end{table}

\begin{figure}[t]
\begin{center}
\includegraphics[width=8cm]{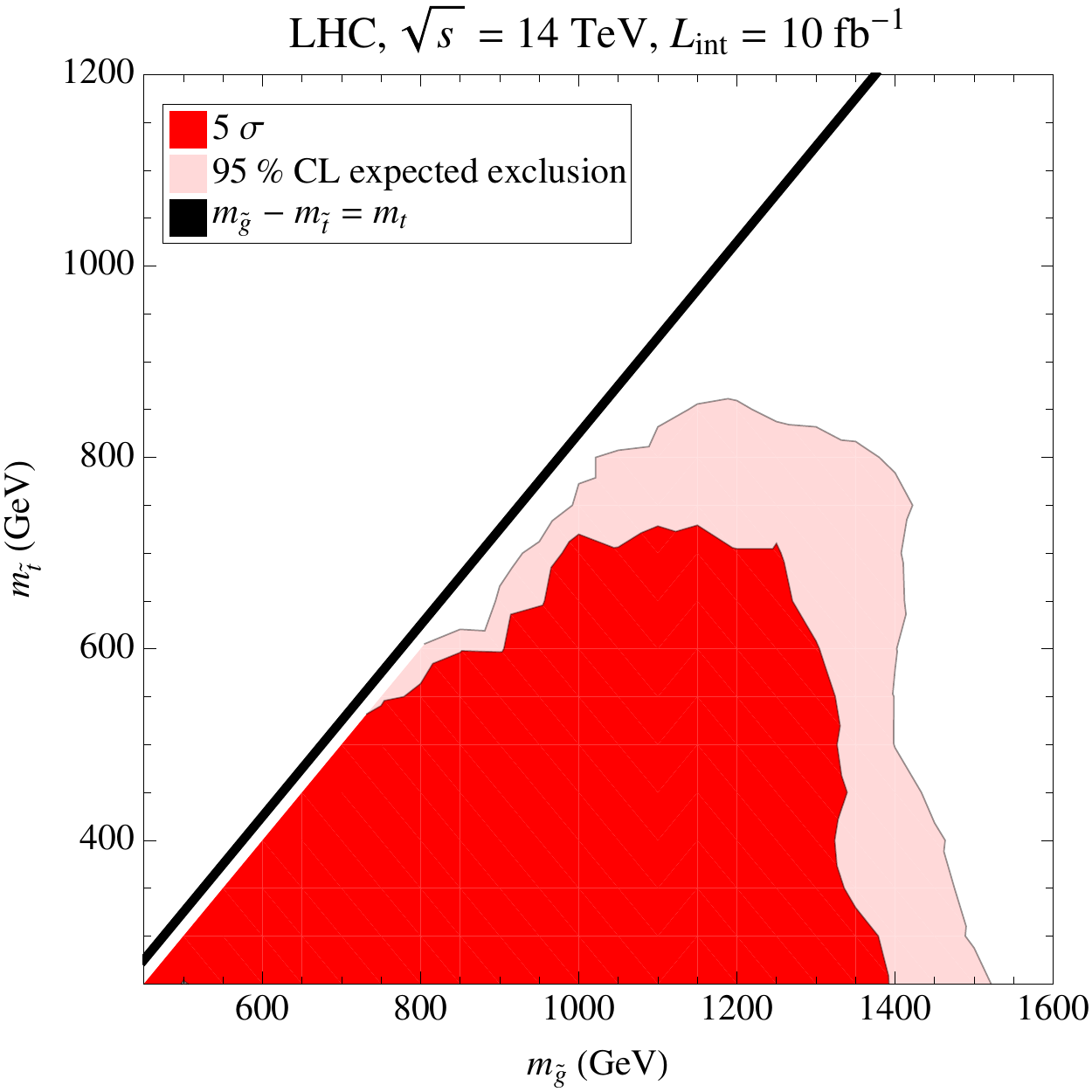}
\caption{The 95\% c.l. expected exclusion and 5-sigma discovery reach of the proposed search at the 14 TeV LHC run with 10 fb$^{-1}$ integrated luminosity.}
\label{fig:reach14}
\end{center}
\vspace{-8mm}
\end{figure}

{\em LHC Sensitivity at $\sqrt{s}=14$ TeV ---} Anticipating higher reach of the search at 14 TeV, we optimize the selection cuts for a benchmark point with higher masses, $(m(\tilde{g}), m(\tilde{t}))=(1200, 600)$ GeV. After again considering all possible combinations of loose and tight tag requirements, we conclude that the optimal strategy in this case is to require three loose tags. We further require $\met\geq 175$ GeV. At the benchmark point, we expect 8.5 signal events to pass these cuts in a data set  of 10 fb$^{-1}$, and with $S/B=27.5$ the expected statistical significance of observation is $6.5$. The reach of a search with these parameters is shown in Fig.~\ref{fig:reach14}. Discovery is possible up to $1.3-1.4$ TeV gluino masses with stops in the $300-700$ GeV mass range. In this case, $S/B\gsim10$ throughout the discovery region.

Given how effective the top tagging technique is in suppressing backgrounds, it is natural to wonder whether, given enough data, a search for gluinos could be conducted with no MET requirement at all. Unfortunately, this is not possible. While the backgrounds studied above are sufficiently suppressed, a new irreducible background, pure QCD events with 4 hard jets ($p_T>100$ GeV), must be included in the absence of a MET cut. The rate for this process is so large ($5.3$ nb at 14 TeV at tree-level) that, even 
including the small mistag probabilities for light jets, it overwhelms the signal. We estimate that the most sensitive search without a MET cut is again one with 3 loose top tags required. For a $300$ fb$^{-1}$ data set, this search is sensitive to the benchmark point at about 4.5 sigma level (statistics-only), but with $S/B\sim 0.1$, systematic errors are probably too large to claim sensitivity.

{\em Discussion ---} Our analysis indicates that using tagged top jets as an additional handle to suppress SM backgrounds in the search for gluino decaying to stops leads to interesting reach, even in the  7 TeV run. In fact, the reach may be even higher than we estimate, since we did not perform a thorough cut optimization for various regions of the model parameter space, instead simply freezing the cuts to values that were found to be near-optimal for a single benchmark point.

While we made several simplifications in this exploratory study, the promising results in our opinion justify a more complete analysis. Most of the outstanding issues concern backgrounds. For irreducible backgrounds, the fixed-order (tree-level) simulations used here should be supplemented with showering and hadronization, although since the jets used in our analysis are required to have rather high $p_T$, we  do not expect qualitative changes. Also, MC samples with higher statistics should be used to fully simulate top-tagging efficiencies on the backgrounds. Reducible backgrounds, which were ignored here, should be studied. The most important one of these is the pure QCD channel, $4j$ at parton level, which has a very high rate even with a 2 or 3 mistagged-jet requirement. The pure-QCD events passing our cuts lie far on the tail of the MET distribution for this channel, where the MET is entirely due to undetected or incorrectly measured jets. Correctly estimating this background would thus be a task for a complete detector simulation or a data-driven approach, which must be performed by the experimental collaborations. It is important to note, however, that large-MET QCD tails affect all SUSY searches at the LHC relying on MET, and in the purely hadronic searches this effect is typically subdominant to the reducible backgrounds once appropriate cuts are applied to eliminate events with MET aligned with one of the jets~\cite{LHC_data}. Similar techniques can be applied in our case. 

{\em Conclusions ---} If SUSY is realized in such a way that stops and sbottoms are the only squarks below the TeV scale, as favored by naturalness and recent negative results from the LHC, top-rich final states are a natural place to search for it. Our results indicate that the techniques to separate top jets from light jets, developed recently with a completely different motivation, can be employed to boost sensitivity of such searches. They can complement other proposed strategies for this scenario~\cite{Kane_tops,BKT}, especially in the heavy gluino region. We encourage the experimental collaborations to incorporate this tool in the upcoming searches.  

{\em Acknowledgments ---} MP is grateful to Michael Peskin for the conversation which led to this project. This work is supported by the U.S. National Science Foundation through grant PHY-0757868 and CAREER award PHY-0844667, and the National Sciences and Engineering Research Council for Canada (NSERC).

\end{document}